\documentclass[a4paper]{jpconf}
\usepackage{graphicx}
\begin{document}
\title{Event by Event fluctuation in $K/\pi$ ratio at RHIC}

\author{Supriya Das (for the STAR Collaboration \footnote{Complete author list can be found at the end of this proceedings})}

\address{Experimental Physics Division, 
Variable Energy Cyclotron Center,\\ 
1/AF Bidhan Nagar, 
Kolkata -700064, 
INDIA}

\ead{supriya@veccal.ernet.in}

\begin{abstract}
We present the preliminary results from our analysis of event by event fluctuation in $K/\pi$ ratio in Au+Au collision at $\sqrt s_{NN}$ = 200 GeV and at 62.4 GeV using STAR detector at RHIC. Two different methods have been used to extract the strength of dynamical fluctuation and to study the centrality dependence of that. The results from the study of energy and centrality dependence of the dynamical fluctuation are presented. From the excitation function it is seen that at two RHIC energies the measure of dynamical fluctuation is constant with values very close to that at 12.3 GeV at SPS. The dynamical fluctuation is found to be positive and decreasing with increasing centrality at RHIC. The results are compared with HIJING model calculation with jets. Results from HIJING are found to be very close to data from central collisions whereas the model over predicts the data for peripheral events. 
\end{abstract}.

\section{Introduction}
A key question of the heavy ion program at the Relativistic Heavy Ion Collider (RHIC) is to understand whether the hot and dense matter produced in the midst of the relativistic heavy ion collisions undergoes a transition to and from a quark gluon plasma (QGP) phase before it hadronizes. The study of fluctuations of specific observables on event by event basis, can provide comprehensive insight in this matter [1-7]. Several recent experimental studies at the SPS and RHIC have focused on the study of fluctuations in relativistic heavy ion collisions. Results from NA49 experiment at SPS \cite{na49prl,guntharqm04} show interesting behavior of both $K/\pi$ ratio as well as the event by event dynamical fluctuation in that quantity. In this paper we present preliminary results from studies on the event by event fluctuation in the $K/\pi$ ratio for Au+Au collisions at $\sqrt s_{NN}$ = 200 GeV and at 62.4 GeV using the STAR detector at RHIC.   

\section{STAR experiment}
STAR experiment at RHIC measures charged hadrons over a wide range of pseudorapidity ($|\eta| < 1$) and azimuthal angle ($-\pi < \phi <\pi$), using the large Time Projection Chamber \cite{tpc}. An overview of the experiment can be found in \cite{starnim}. The kaons and pions are identified based on their relative energy deposition (dE/dx) in the TPC within the momentum range of 100 MeV - 600 MeV. This wide acceptance of the detector enhances the possibility to study event by event physics. Specially the wide momentum range available for particle identification enables to study event by event fluctuation in $K/\pi$ ratio.

\section{Analysis}
We have used the dataset from RHIC run II for Au+Au 200 GeV and RHIC run IV for Au+Au at 62.4 GeV. Around 300K minimum bias events, corresponding to 0 - 80\% of Au+Au hadronic interaction cross-section, are analysed for this work. In minimum bias mode the events were triggered by two ZDCs and a minimum signal at the CTB. Only events for which the z-vertex is within $\pm 25$ cm of the center of the TPC are accepted. The global tracks with at least 15 hits in TPC and z component of Distant of Closest Approach (DCA) of the track with the primary vertex is less than 3 cm are used in this analysis.   
Two separate methods of analysis are used, one to obtain the overall measure of the dynamical fluctuation in central events and the other one to obtain the centrality dependence of the dynamical fluctuation.

\subsection{Dynamical fluctuation and its excitation function}
The kaons and pions in each event are identified from their relative energy deposition in TPC. Then the ratio of multiplicities of kaons and pions is obtained on event by event basis. The width ($\sigma_{data}$) of the distribution of this ratio for sufficiently large number of events, gives the fluctuation in the ratio for real data. In order to isolate the strength of non-statistical fluctuation, we need to establish a reference which tells us the contribution to fluctuation from finite number of particles detected in the detector, experimental resolution in dE/dx but not containing any further correlations between the particles. This statistical limit corresponding to uncorrelated kaon and pion production, is estimated based on the ratio of kaon and pion multiplicities obtained from pseudo events constructed using a mixed event technique ($\sigma_{mixed}$). This mixing of the events has been done picking up one track from every event randomly to construct one mixed event while the multiplicity of the mixed events follow that of the real events. The dynamical fluctuation in the quantity has been obtained by the formula

\begin{equation}
\sigma_{dyn} = \sqrt{\sigma_{data}^2 -\sigma_{mixed}^2}     
\end{equation}  

This approach was used by NA49 to measure the strength of dynamical fluctuation for central events at various SPS energies. We have therefore used this to compare our results with those from SPS.
 
\subsection{Results}

\begin{figure}[h]
\begin{minipage}{18pc}
\includegraphics[width=18pc]{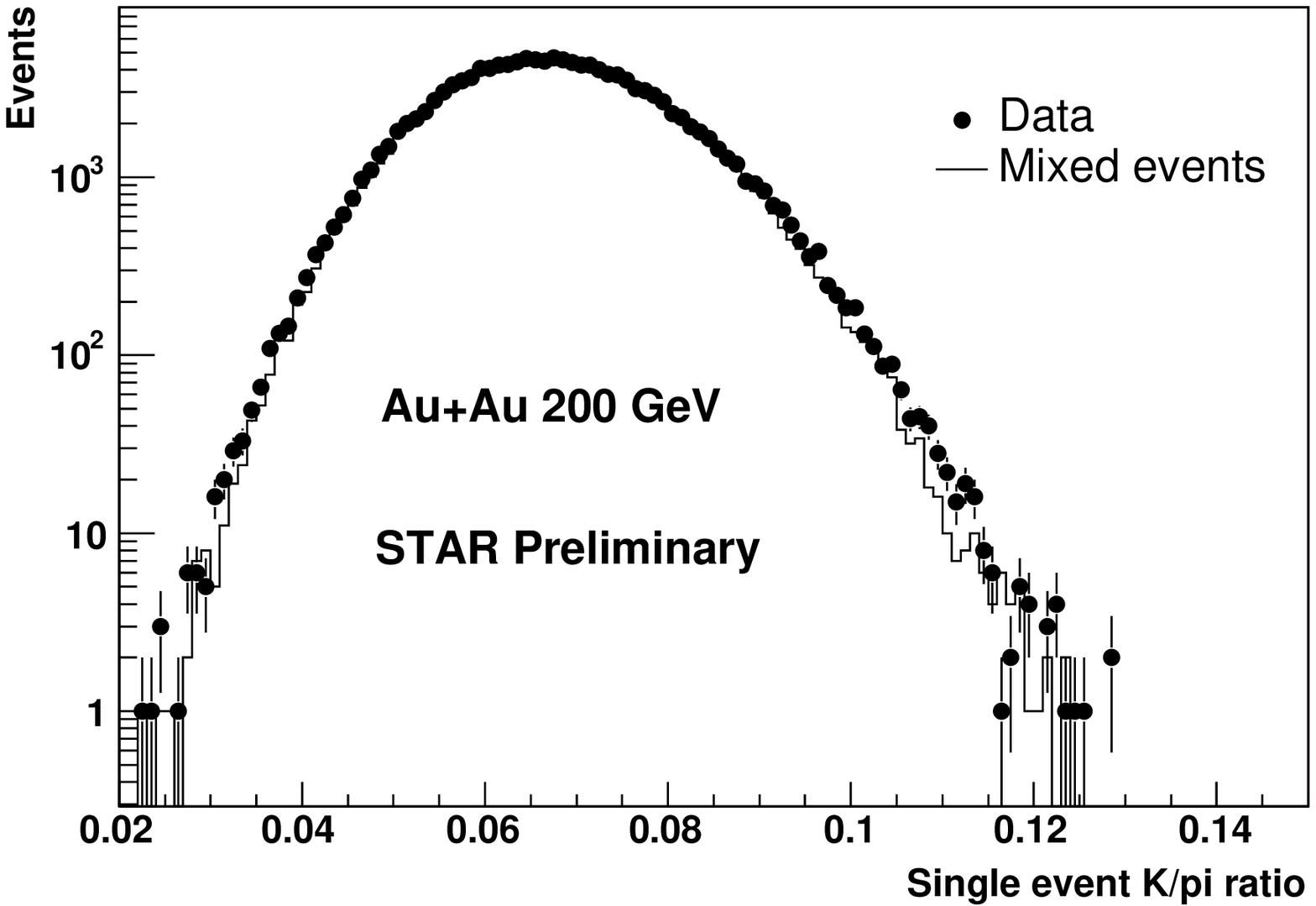}
\caption{\label{sigma200}Distribution of $K/\pi$ ratio from data and mixed events for Au+Au at 200 GeV}
\end{minipage}\hspace{2pc}%
\begin{minipage}{18pc}
\includegraphics[width=18pc]{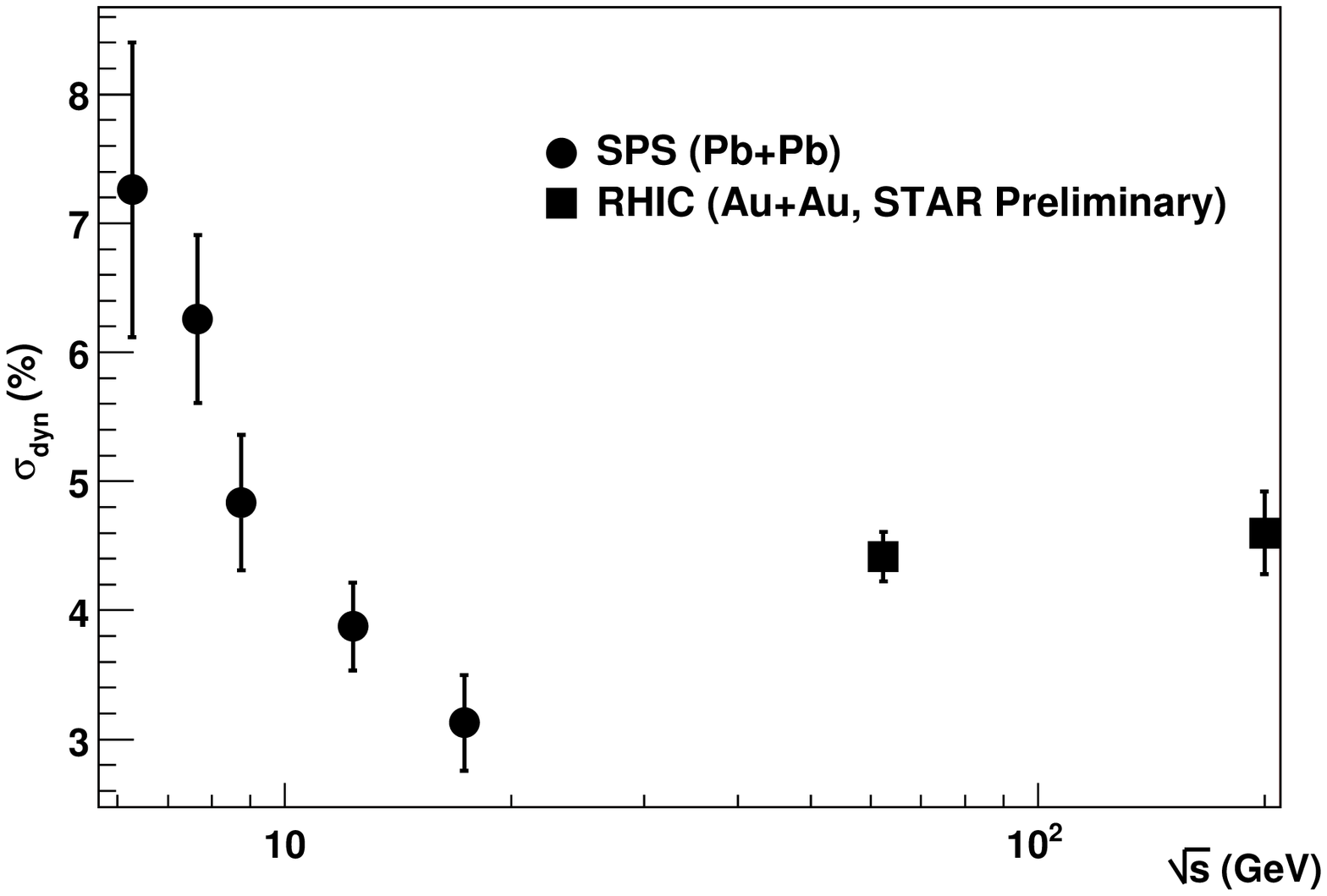}
\caption{\label{sigmaexcitation}Excitation function for $\sigma_{dyn}$}
\end{minipage} 
\end{figure}

Fig.~\ref{sigma200} shows the distribution of $K/\pi$ ratio for top 5\% central events in Au+Au collisions at 200 GeV. The distributions obtained from real and mixed events are superimposed to help comparing the widths of the distributions. The values of $\sigma_{dyn}$ in different cases are tabulated in the table below.

In Fig.~\ref{sigmaexcitation} the energy dependence (excitation function) of $\sigma_{dyn}$ is plotted. The data points up to 17 GeV are obtained from NA49 collaboration \cite{guntharqm04}. At two RHIC energies, the measure of dynamical fluctuation appears to be constant having values close to SPS data at 12.3 GeV.

\begin{table}[h]
\caption{\label{sigmatable}Values of $\sigma_{dyn}$}
\begin{center}
\begin{tabular}{llllll}
\br
Experiment&System&Ratio type&$\sigma_{data}$&$\sigma_{mixed}$&$\sigma_{dyn}$\\
\mr
NA49&Pb+Pb at 17.3 GeV&$K/\pi$& - & - &3.12($\pm$0.37)\%\\
STAR&Au+Au at 200 GeV&$K/\pi$&17.78\%&17.23\%&4.6($\pm$0.38)\%\\
STAR&Au+Au at 200 GeV&$K^+/\pi^+$&24.29\%&24.10\%&3.06($\pm$0.88)\%\\
STAR&Au+Au at 200 GeV&$K^-/\pi^-$&24.81\%&24.55\%&3.61($\pm$0.67)\%\\
STAR&Au+Au at 62.4 GeV&$K/\pi$&20.7\%&20.3\%&4.44($\pm$0.16)\%\\
\br
\end{tabular}
\end{center}
\end{table}

\subsection{Centrality dependence of dynamical fluctuation}
To study the centrality dependence of the dynamical fluctuation we have used the variance $\nu_{dyn}$, first suggested in Ref. \cite{nudyn0}, which is derived from single and two-particle distribution functions. In this method the dynamical fluctuation has been expressed as the contribution from three terms representing the correlation strength of kaons, pions and the cross correlation between them \cite{nudyn1} :

\begin{equation}
R=R_{KK}+R_{\pi \pi}-2R_{K\pi}
\end{equation}

This approach has been used in several cases where the fluctuation of the ratio of two quantities is measured \cite{nudyn2}.

Experimentally this quantity is obtained from the following relation:

\begin{equation}
 \nu_{dyn,K/\pi}(M)=\frac{\langle K(K-1)\rangle_M}{\langle K \rangle^2_M}+\frac{\langle \pi(\pi-1) \rangle_M}{\langle \pi \rangle^2_M}-2\frac{\langle K\pi \rangle_M}{\langle K \rangle \langle \pi \rangle_M}
\end{equation}

The notation $\langle O \rangle_M$ is used to indicate the average of the quantity $O$ for all events with uncorrected multiplicity $M$ of charged hadrons within pseudorapidity range $|\eta|<0.5$. Centrality is defined dividing this whole range of multiplicity in nine contiguous bins.

\subsection{Results}

\begin{figure}[h]
\includegraphics[width=18pc]{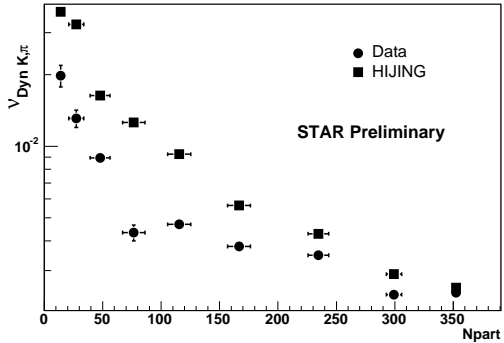}\hspace{2pc}%
\begin{minipage}[b]{18pc}\caption{\label{k2pinudyn}Centrality dependence of $\nu_{dyn,K/\pi}$ for Au+Au collisions at $\sqrt s_{NN}$ = 200 GeV.}
\end{minipage}
\end{figure}

Fig. ~\ref{k2pinudyn} shows the variation of $\nu_{dyn,K/\pi}$ with centrality where along X-axis plotted is $N_{part}$ which is the number of participants in Au+Au collisions calculated by optical glauber model calculation. Error on the data points correspond to statistical errors only. Results from HIJING\cite{hijing} calculation is shown to compare with the data. 10K HIJING events for each centrality bin have been used for this purpose.
 
\section{Summary}

 We have presented the preliminary results of the event by event fluctuation in $K/\pi$ ratio measured at RHIC using the STAR detector. Two different methods were used to extract the dynamical fluctuation. In one method the distribution of $K/\pi$ ratio from data was compared with that obtained from mixed events which provides the statistical limit of this fluctuation, corresponding to uncorrelated kaon and pion production. The results from this method shows the dynamical fluctuation in $K/\pi$ for top 5\% central Au+Au at 200 GeV as 4.6\%, for $K^+/\pi^+$ as 3.06\%, for $K^-/\pi^-$ as 3.61\% and for $K/\pi$ for top 5\% central events for Au+Au at 62.4 GeV as 4.44\%. The excitation function of dynamical fluctuation has also been studied combining RHIC data with data from SPS. At two RHIC energies, the measure of dynamical fluctuation appears to be constant having values closers to SPS data at 12.3 GeV. Another method based on correlation studies were used to study the centrality dependence of this dynamical fluctuation was used. The result from this shows the dynamical fluctuation for Au+Au 200 GeV is positive and decreases with increasing centrality. These results were compared with those obtained from HIJING and the model calculation results are very close for the central collisions whereas for peripheral events the model overpredicts the data.

\section{Acknowledgments}
We thank the RHIC Operations Group and RCF at BNL, and the
NERSC Center at LBNL for their support. This work was supported
in part by the HENP Divisions of the Office of Science of the U.S.
DOE; the U.S. NSF; the BMBF of Germany; IN2P3, RA, RPL, and
EMN of France; EPSRC of the United Kingdom; FAPESP of Brazil;
the Russian Ministry of Science and Technology; the Ministry of
Education and the NNSFC of China; Grant Agency of the Czech Republic,
FOM of the Netherlands, DAE, DST, and CSIR of the Government
of India; Swiss NSF; the Polish State Committee for Scientific
Research; and the STAA of Slovakia.

\end{document}